\documentclass[prb,nobalancelastpage,twocolumn,nofootinbib,superscriptaddress,showpacs]{revtex4-1}

\usepackage{physics}
\usepackage{graphicx}% Include figure files
\usepackage{bm}
\usepackage{bbm}% bold math
\usepackage[bb=boondox]{mathalfa}
\usepackage{nicefrac}
\usepackage{subcaption}
\usepackage{verbatim}
\usepackage{scrextend}
\usepackage{tikz}
\usepackage{bbold}

\usepackage[justification=justified,
   format=plain]{caption}

\usepackage{amsfonts}
\usepackage{amssymb}
\usepackage{amsmath}
\usepackage{epsfig}
\usepackage{enumerate}
\usepackage{multirow}

\usepackage{caption}
\usetikzlibrary{positioning}
\usetikzlibrary{fit}
\usetikzlibrary{calc,shadings}
\usetikzlibrary{patterns}

\begin{document}

\title{Behavior of l-bits near the many-body localization transition}% Force line breaks with \\

\author{Abishek K. Kulshreshtha}
\author{Arijeet Pal}%
\author{Thorsten B. Wahl}
\author{Steven H. Simon}
\affiliation{%
 Rudolf Peierls Centre for Theoretical Physics, Oxford, 1 Keble Road, OX1 3NP, United Kingdom.
}%

\date{\today}% It is always \today, today,

\begin{abstract}
Eigenstates of fully many-body localized (FMBL) systems are described by quasilocal operators $\tau_i^z$ (l-bits), which are conserved exactly under Hamiltonian time evolution. The algebra of the operators $\tau_i^z$ and $\tau_i^x$ associated with l-bits ($\boldsymbol{\tau}_i$) completely defines the eigenstates and the matrix elements of local operators between eigenstates at all energies. We develop a non-perturbative construction of the full set of l-bit algebras in the many-body localized phase for the canonical model of MBL. Our algorithm to construct the Pauli-algebra of l-bits combines exact diagonalization and a tensor network algorithm developed for efficient diagonalization of large FMBL Hamiltonians. The distribution of localization lengths of the l-bits is evaluated in the MBL phase and used to characterize the MBL-to-thermal transition. %Both the average and typical localization lengths ($\bar{l}$) diverge as $\bar{l} \sim (h-h_c)^{-\nu}$ for $\nu_{\text{avg}} \approx 0.56$ and $\nu_{\text{typ}} \approx 0.44$ for critical disorder strength $h_c \approx 2 - 2.9$. We compare the tails of the probability distribution to exponential and power law forms. Deep in the localized phase the exponential distribution is a better fit while close to the transition due to finite-size effects the two distributions are indistinguishable.           
\end{abstract}

\maketitle

%\tableofcontents

%\section{\label{sec:intro}Introduction}

\emph{Introduction:} Dynamics of thermalization in closed, interacting quantum systems is a phenomenon of fundamental importance which has received considerable attention in the past decade \cite{Polkovnikov:2011RMP}. Although states of matter at thermal equilibrium are widespread, the phenomenon of MBL has provided a novel paradigm for the breakdown of thermalization in quantum systems \cite{basko2006metal, gornyi2005interacting, pal2010mb, oganesyan2007localization, nandkishore2015many}. MBL has now been realized in several experiments using cold atoms and trapped ions and shown to be a robust phase \cite{ Schreiber842, Smith_MBL}. MBL as a quantum phase of matter raises several exciting possibilities for realizing topological order in excited states and preserving quantum information \cite{Huse2013LPQO, Chandran2014SPT, bahri2015localization}. Although MBL has been firmly established in one dimension \cite{imbrie2016many, Imbrie2016MBL}, several questions related to its instability to thermalization at weaker disorder \cite{VHA_MBLTransition, Potter:2015ab, Khemani2017MBLT} and existence in higher dimensions remain hotly debated \cite{chandran2016higherD, Roeck2016stability, Wahl2017TNS_2D}.          

Many-body eigenstates of thermal systems are exponentially complex. On the other hand, for MBL in one dimensional models with bounded local Hilbert spaces \cite{FMBLfootnote}, an efficient description emerges when the entire spectrum is localized, due to an extensive set of local conservation laws given by the operators, $\tau_i^z$, known as \emph{l-bits} \cite{huse2014phenomenology, serbyn2013local, ros2015integrals}. Approximate $\tau_i^z$ operators represent the entire spectrum with exponential accuracy in terms of the quantum numbers of these operators, which scale only linearly with the size of the system. Such a structure of the eigenstates also implies the existence of quasi-local operators $\tau_i^x$ which produce transitions between two particular many-body eigenstates. For a spin-$1/2$ stystem, the $\tau_i^z$ and $\tau_i^x$ operators satisfy the Pauli spin algebra where the full many-body Hamiltonian is diagonal in the $\tau_i^z$ basis, and $\tau_i^x$ operators characterize the matrix elements between the eigenstates at all energies. Therefore, l-bits are analogoues of bare spins, yet describe an interacting system over a range of parameters.      

For finite-size MBL systems the l-bits can be constructed approximately using local unitary transformations \cite{pekker2016fixed, Rademaker2016LIOM, Rademaker2017}. The l-bits constructed in this perturbative manner only commute approximately with the Hamiltonian. The methods accessing the exact $\tau_i^z$ operators by studying the infinite time limit are not able to construct the algebra of the l-bit operators \cite{chandran2015constructing}. In this article we develop a non-perturbative construction of the set of $\tau_i^z$ and $\tau_i^x$ operators representing the l-bits in the MBL phase using a combination of tensor network methods \cite{Wahl2017PRX, Pollmann2016TNS, Pekker2017MPO, Chandran2015STN} and exact diagonalization. In contrast to prior work, the l-bit algebras that we construct are exact and exponentially localized \cite{pekker2016fixed, Rademaker2016LIOM, Rademaker2017}, i.e. the commutator of the set l-bit operators $\tau_i^z$ with the Hamiltonian is strictly zero. Our construction of l-bits allows us to study the behavior of the conserved quantities over a range of disorder strengths, even in the vicinity of the MBL-to-thermal transition.

By performing a finite-size scaling collapse of the l-bit localization length, the critical disorder strength and the correlation length exponent are extracted. For finite-size systems, the divergence of the localization length is cut-off by the system size. The finite-size scaling function also provides an estimate of the cross-over scale between the thermal and the quantum critical regimes. We characterize the distribution of localization lengths of the l-bits as a function of disorder. In the localized phase, the distribution has exponential tails which shows that the l-bits with large localization lengths are rare. On approaching the transition into the thermal phase, the distribution becomes heavy-tailed with significant weight at localization lengths comparable to the system size. The heavy-tails can be fitted to a power law. Due to finite-size effects, an exponential fit is also feasible. Thus, the l-bits with large localization lengths are no longer rare and can be destabilized by local perturbations which produce long-range resonances \cite{imbrie2016many, Imbrie2016MBL}. %We study the average and typical localization lengths of the l-bits as a function of disorder, which appear to diverge close to the expected location of the MBL transition, albeit with slightly different exponents. 

%In Section \ref{sec:model}, we describe the model used and the salient features of l-bit phenomenology in MBL systems. In Section \ref{sec:construction} we provide the detailed methodology to self-consistently construct the algebra of exact l-bits using the map of eigenstates from the optimized tensor network represenation of the diagonalizing unitary operator \cite{wahl_pal_simon}. We further define a metric to extract the localization length of the l-bit operators. It is based on the weight of an operator at a site, which decays exponentially with the distance from the localization center. We characterize the distribution of localization lengths on approaching the transition into the thermal phase. In Section \ref{sec:concl} we summarize the results and discuss future possibilities for these methods.

\emph{Model and l-bit phenomenology:} We work with the one-dimensional XXZ spin chain in a random magnetic field, with the Hamiltonian 
\begin{equation} 
\label{eq:main_ham} 
H = \sum_{i = 1}^{L-1} \mathbf{S}_i \cdot \mathbf{S}_{i+1} + \sum_{i = 1}^L h_i S^z_i, 
\end{equation}
where $\mathbf{S}_i = \frac{1}{2} \boldsymbol{\sigma}_i$ and each $h_i$ is drawn randomly from a uniform distribution $[-h,h]$. The phase diagram of this model is well-studied using exact diagonalization and has a phase transition from the thermal into the full MBL phase at $h_c \approx 3.5$ \cite{pal2010mb, luitz2015many}. %This model is commonly used to study MBL systems \cite{arijeet_thesis}. %The value $h$ will hereafter be referred to as the disorder strength. 

The set of Pauli operators $\{\boldsymbol{\sigma}_i\}$ define the physical bits (`p-bit' operators) which act on a local 2-dimensional Hilbert space. %These p-bits allow us to flip and measure spin states of the system, described by a string of length $n$ with each value $\pm 1$. 
At disorder strengths much larger than $h_c$, due to MBL of the full spectrum, the p-bits can be unitarily transformed into localized bits (`l-bit' operators). Each l-bit operator $\boldsymbol{\tau}_i$ is derived from the corresponding p-bit on site $i$, has weights which decay exponentially with the distance from site $i$. The hallmark of the l-bits is the \emph{exact} commutativity of all $\tau_i^z$ with the Hamiltonian for a finite-size system. These operators are constructed according to $\tau^z_i = U \sigma^z_i U^\dagger$, where $U$ is an operator that diagonalizes the Hamiltonian and preserves the local structure. Notably, the same $U$ that transforms p-bits to l-bits diagonalizes the Hamiltonian. This feature is not known to exist generically outside of the MBL phase. 

%In the $\tau_i^z$ basis the Hamiltonian is diagonal can be written 
%\begin{equation} 
%H = \sum^n_{i = 1} J_i \tau^z_i + \sum^n_{i,j = 1} J_{ij} \tau^z_i \tau^z_j + \sum^n_{i,j,k = 1} J_{ijk} \tau^z_i \tau^z_j \tau^z_k + \ldots, 
%\end{equation} 
%where the coefficients $J_{ijk \ldots}$ decay exponentially with the largest distance between spins $|i-j|$ \cite{ imbrie2016many, Imbrie2016MBL, huse2014phenomenology, serbyn2013local, ros2015integrals}.

%L-bit operators obey the following commutation relations: \begin{equation} \label{eq:comrules} \left[ H,\tau^z_i \right] = \left[\tau^z_i,\tau^z_j\right] = 0. \end{equation}

Note that matrices which diagonalize the Hamiltonian are non-unique. For example, the columns of any matrix diagonalizing the Hamiltonian can be permuted to form another matrix which also diagonalizes the Hamiltonian, yet these permutations affect the the locality of the  $\tau^z_i$ operators. Therefore, not simply any choice of matrix $U$ diagonalizing the Hamiltonian will successfully construct the most local set of $\tau^z_i$; the permutations are constrained to preserve the local structure of the full unitary.

%The eigenstates of both $H$ and $\tau^z_i$ are given by the columns of $U$. The $\tau$ operators, as unitary transformations on the Pauli $\sigma$ operators, are also traceless. Additionally, unitary transformation preserves the eigenvalues of $\sigma$ as well: $2^{n-1}$ of the eigenvalues of $\tau$ are $+1$ and $2^{n-1}$ are $-1$. 

Due to the emergent integrability in the MBL phase, each eigenstate can be labeled by the set of eigenvalues $l_i = \pm 1$ of $\{\tau^z_i\}$. An eigenstate $\ket{\alpha}$ comes with its corresponding ordered string of $\{l^\alpha_i\},$ of length $n$ that are either $+1$ or $-1$. We define the $j$-partner of an eigenstate to be the eigenstate obtained by flipping the $jth$ l-bit. We assign the $j$-partner of $\ket{\alpha}$ as $\ket{\beta_{j,\alpha}}$ and they are said to be \textit{paired} on site $j$. The structure of the partnering of eigenstates may not be unique and provides a representation of the l-bit operators, which are constructed from the eigenstates as given in Eqs. \ref{eq:lbitsx}-\ref{eq:lbitsz}:
\begin{align}  \label{eq:lbitsx}
\tau^x_i &= \sum_{\alpha} \ket{\alpha} \bra{\beta_{i,\alpha}} \\ \label{eq:lbitsy}
\tau^y_i &= -i \sum_{\alpha} l^\alpha_i \ket{\alpha} \bra{\beta_{i,\alpha}} \\
\label{eq:lbitsz}
\tau^z_i &= \sum_{\alpha} l^\alpha_i \ket{\alpha} \bra{\alpha}.
\end{align}
In this construction, $\tau^x_i$ plays the role of a bit flip operator, similar to the $\sigma^x_i$ Pauli matrix. For example, $\tau^x_2$ will flip the eigenstate $\{ + + + \ldots \}$ to $\{+ - + \ldots \}$. The network of allowed partner eigenstates are tightly constrained by the algebraic structure of the $\tau_i^x$ operators. 

The action of the l-bit operators $\tau_i^z$ and $\tau_i^x$ on the eigenstates is very simple. When $\tau^z_i$ acts on an eigenstate $\ket{\alpha}$, the eigenstate is returned with a sign $\pm 1$ to match $l^\alpha_i$. There are no off-diagonal matrix elements in the eigenstate basis. When $\tau^x_i$ acts on an eigenstate, it produces a transition to an eigenstate with the l-bit eigenvalue on site $i$ being flipped. In this sense, the l-bits can simply be thought of as ``dressed" p-bits and are related to them by a sequence of local unitary transformations, with weight exponentially decaying with distance from the l-bit's localization center. 

\emph{Construction of exact l-bit algebras:} In order to construct l-bit operators as described in Equations (\ref{eq:lbitsx})-(\ref{eq:lbitsz}), we must find an eigenstate pairing structure (or a configuration of $j$-partners) that creates quasi-local operators. The tensor network approach developed in \cite{Wahl2017PRX} provides an efficient method to approximate the unitary $U$ which transforms the Hamiltonian into a predominantly diagonal basis for an MBL system. The unitary operator $U$ provides a natural l-bit structure of the approximate eigenstates and their pairing structure. By matching exact eigenstates to these approximate ones, we can find a pairing scheme that produces quasi-local operators that exactly commute with the Hamiltonian \cite{SupplementA}.
\begin{figure}
\includegraphics[width=0.8\linewidth]{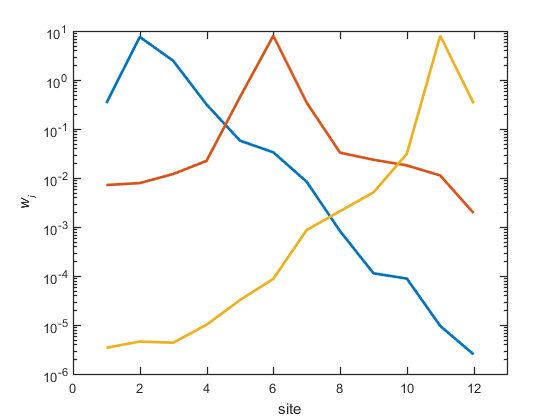}
\caption[width=\linewidth]{ Weight functions $w_i (\tau_z)$ (see eq. \ref{eq:weight}) showing exponential decay with position for three $\tau_z$ operators. Operators shown are for a single disorder realization at $h=8$. The localization lengths (defined in the main text) are $\xi_z = 0.67$, $0.86$, and $0.76$ respectively. The r-values of the exponentially decaying fits on the weight functions are  $R^2 = 0.99$, $0.85$, and $0.89$ respectively.}\label{fig:1}
\end{figure}

In order to characterize the localization properties of our exact l-bit operators, we propose a measure of localization length of an operator. Any operator $\hat{O}$ can generally be written in the form $\hat{O} = \sum_{\gamma \in \{0,x,y,z\}}  A^\gamma_{\bar{i}} \otimes \sigma^\gamma_i$, 
for any site $i$. Here, $\sigma^0$ is the identity operator and the matrices $A^\gamma_{\bar{i}}$ act on all sites that are not $i$. 

An operator is local to a site $i$ if for $j \neq i$, $A^x_{\bar{j}} = A^y_{\bar{j}} = A^z_{\bar{j}} = \mathbb{0}$. Thus, a scalar value for the weight of an operator at site $i$ is given by: $\Tr [{A^x_{\bar{i}}}^2 + {A^y_{\bar{i}}}^2 + {A^z_{\bar{i}}}^2]$. For a maximally local operator located on site $i$, $w_j = 0$ for all $j \neq i$. For l-bit operators, where the weight of the operator is expected to decay exponentially, the localization length is finite.
We define 
\begin{align} 
w_i (\hat{O}) &\equiv 8 \Tr [{A^x_{\bar{i}}}^2 + {A^y_{\bar{i}}}^2 + {A^z_{\bar{i}}}^2] \nonumber \\ \label{eq:weight} 
&= \sum_{\gamma \in \{x,y,z\}} \Tr [(\hat{O} - \sigma^\gamma_i \hat{O} \sigma^\gamma_i)^2]. 
\end{align} 
For a quasi-local operator, the weight of an operator centered on site $i$ should decay as $e^{-|i-j|/\xi_z}$. In fact, Figure \ref{fig:1} shows the exponential decay of three $\tau^z$ operators for a single disorder realization at $h = 8$ for system size $L=12$. We define a localization length as the value of $l$ that produces the best fit curve for the weight of the operator \cite{SupplementB}.
%\begin{equation}
%l = \xi_b \label{eq:loclength}
%\end{equation}
%, where $\xi_b$ is the value of $\xi$ that produces the best fit curve for the weight of the operator. 
This best fit curve is calculated on the weight function of the operator from the peak weight to the furthest boundary.% \begin{equation} \label{eq:loc_length} l^2 \equiv \frac{\displaystyle \sum_{i = 1}^n \sum_{j = 1}^n w_i w_j (i-j)^2}{\displaystyle \sum_{i = 1}^n \sum_{j = 1}^n w_i w_j} . \end{equation}\begin{equation} \label{eq:loc_length} l^2 \equiv \left. ^{\displaystyle \sum_{i = 1}^n \sum_{j = 1}^n w_i w_j (i-j)^2} \middle/ _{\displaystyle \sum_{i = 1}^n \sum_{j = 1}^n w_i w_j} \right. . \end{equation}

It should be noted that this method may not necessarily find the most local integrals of motion for a given system. Rather, the algorithm finds highly localized operators that commute with the Hamiltonian and with one another; this does not preclude the possibility of a pairing structure that yields more local operators that commute with the Hamiltonian and with one another.

\emph{Distribution of localization lengths and MBL transition:} 
A rigorous proof of the l-bit framework to describe MBL is valid only deep in the localized phase \cite{imbrie2016many, Imbrie2016MBL}. The lack of systematic construction procedures of exact l-bits has left the question of their existence and their statistical properties close to the MBL transition largely in the realm of heuristic arguments. For instance, how do the statistics of the length scale of the operators vary in vicinity of the transition? Deep in the MBL phase, the exponential rarity of l-bits with large localization lengths is crucial for their stability to perturbations on the Hamiltonian. On approaching the transition, the increased likelihood of l-bits with large localization lengths may render these operators unstable raising the possibility of alternative effective description of localization near the transition \cite{chandran2016higherD}. An understanding of the properties of l-bits at lower disorder is crucial for developing a theory of the transition. The method developed here allows us to probe the properties of the length scale of l-bit operators which are always exactly conserved by construction.  
   
\begin{figure}
	\includegraphics[width = 0.8\linewidth]{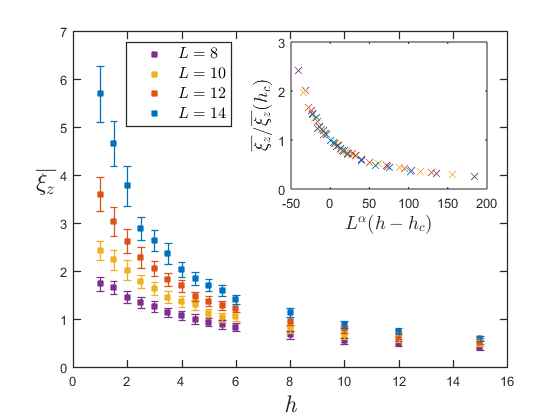}
	\caption{\label{fig:median}Plot of mean localization length ($\overline{\xi}_z$) of $\tau^z_i$ operators constructed on systems of varying lengths ($L$) and disorders ($h$). Error bars shown indicate the standard deviation of the distribution of localization lengths for each length and disorder. In the inset, a scaling collapse is presented. This collapse is achieved through values $\nu =1.1$ and $h_{\text{crit}} = 3.5$. The values of $\nu$ and $h_{\text{crit}}$ were chosen by optimizing a polynomial fit on $\overline{\xi}_z/\overline{\xi}_z(h_{\text{crit}})$}.
\end{figure}

Using approximately 130 realizations for each disorder strength at four different system sizes ($L=8,10,12,14$), we find a systematic increase in the average localization length of the $\tau_i^z$ operators with decreasing $h$. Deep in the localized phase ($h \sim 8$) the localization length is of the order of, or even less than a single lattice spacing. For systems of size $L=14$, at $h=5$ the average localization length $\bar{\xi}_z = 1.71$ and well into the MBL phase $\bar{\xi}_z = 0.60$ at $h=15$. As the Hamiltonian approaches the transition into the thermal phase at lower disorder strengths, the average localization length of the $\tau_i^z$ operator increases for all system sizes, as indicated in Figure \ref{fig:median}. We find that for systems of size $L=14$, the mean localization length $\bar{\xi}_z = 2.63$ for $h=3$.  %\textcolor{blue}{(AP: I have changed the notation a bit, $l_z \rightarrow \xi_z$, $\alpha \rightarrow 1/\nu$ for the argument of $F(x)$ and its algebraic form. This is in order to conform to the more conventional choice of variables and avoid ambiguity with other terms in the text. The figures have to be adjusted accordingly.)}

\begin{figure}
	\includegraphics[width = 0.8\linewidth]{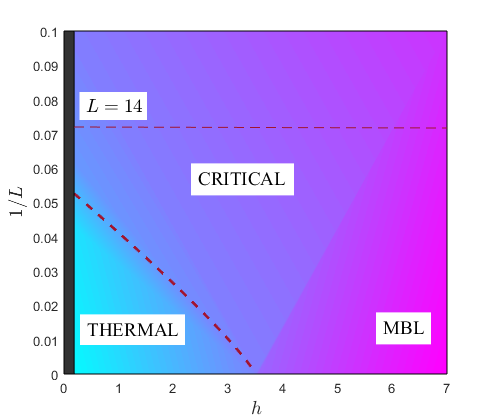}
	\caption{\label{fig:phase}A schematic diagram showing the phases of the disordered XXZ Heisenberg spin chain and the quantum critical crossover regime as a function of disorder $h$ and inverse system size $L$. The bold dashed line represents the crossover scale  ($h_{\text{cross}}$) between the thermal and the quantum critical regime. An estimate of $h_{\text{cross}}$ given by Eq. \ref{eq:crossover} in the main text. Note that the region where $h$ is close to zero is purposely obscured; at this level of disorder, effects due to integrability are likely to take precedence. Note that for the system sizes we have access to, the average localization still remains much smaller than L which is an indication that we are in the quantum critical regime. }%Note further that according to this diagram, our numerics do not include large enough system sizes to see a completely thermal phase even at our lowest disorder of $h=1$. This may explain why we fail to see a complete divergence in the localization length of $l$-bits, even at $h=1$.}
\end{figure}

We then conduct a scaling collapse for the function $\overline{\xi}_z/\overline{\xi}_z(h_{\text{crit}}) = F[L^{1/\nu} (h-h_{\text{crit}})]$, also shown in Figure \ref{fig:median}. Optimal values for the fit are found at $h_{\text{crit}} = 3.5$ and $1/\nu = 0.9$ according to a least-square fit. Further, the function $F$ is shown to be of the form $F(x) \propto (x+x_{\infty})^{-\alpha}$ where $x_{\infty} = 48$ and $\alpha=0.55$. We do not find a significant difference between the scaling collapse of the mean and typical localization lengths. The correlation length exponent $\nu$ is smaller than the finite-size scaling bounds on $\nu$ provided by the Harris criterion for the MBL transition \cite{Chandran:2015ab}, much like other exact diagonalization studies. The scaling function gives an insight into the thermal to quantum critical crossover regime. The fact that there are l-bits with localization length $\xi_z \ll L$ around the critical disorder strength suggests that the system is in the crossover regime for these system sizes. The form of the scaling function $F(x)$ provides a relationship between system size and disorder strength,  $h_{\text{cross}}(L)$,  which describes the regime at which the l-bit localization length diverges for the disordered XXZ Heisenberg chain. This is shown the schematic phase diagram in Figure \ref{fig:phase} which is consistent with the picture of the transition given in \cite{khemani2016critical}. The scaling collapse gives an estimate of this crossover scale on the thermal side of the transition, 
: \begin{equation} 
h_{\text{cross}} (L) = 3.5 - \frac{48}{L^{0.9}}. 
\label{eq:crossover}
\end{equation} 
The crossover between the critical and MBL regimes, which is expected in equilibrium second order transition, is not visible in this quantity. A study of the l-bits with large localization lengths near the transition can contain information of the resonant cluster responsible for the entire crossover regime as seen in phenomenological renormalization group studies \cite{VHA_MBLTransition, Potter:2015ab}.

\begin{figure}
\begin{minipage}{0.8\linewidth}
%\begin{subfigure}
\includegraphics[width=0.85\linewidth]{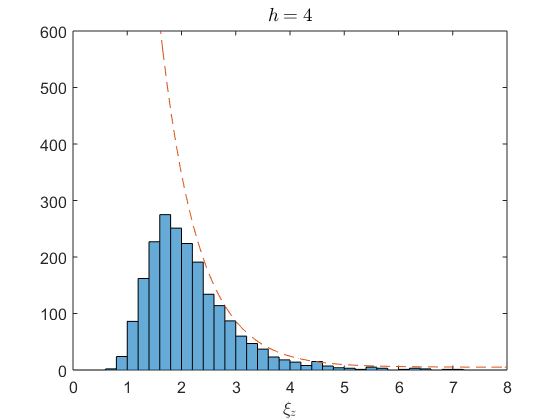}
\subcaption{\label{fig:dist4_6}}
%\caption{\label{fig:dist4_6}}
%\end{subfigure}
\end{minipage}
%\begin{subfigure}
\begin{minipage}{0.8\linewidth}
\includegraphics[width=0.85\linewidth]{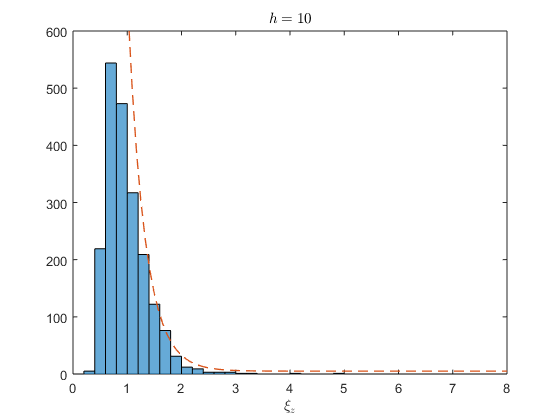}
\subcaption{\label{fig:dist8_10}}
%\caption{\label{fig:dist8_10}}
\end{minipage}
%\label{fig:dist8_10}
%\end{subfigure}
\caption[width=\linewidth]{Histograms of localization lengths of $\tau_i^z$ at system size $L=14$. The histograms show that with increasing disorder strength, the proportion of operators with localization length $\xi_z < 1$ increases. The decay of the tail of each histogram is shown as a dashed line. An exponential decay fit ($P(\xi_z) \propto e^{- \xi_z/\Theta}$, with $\Theta$ the decay-coefficient) was attempted on each graph from the peak of the distribution to the edge. Weaker disorder systems, here $h=4$, fit to exponential decay models with decay coefficient $\Theta= 0.70$ and error $R^2 = 0.98$. They fit with approximately equal quality to a power law ($P(\xi_z) \propto \xi_z^{-\eta}$) with decay coefficient $\eta = 6.5$ and error $R^2 = .97$. Stronger disorder systems, here $h=10$, fit most strongly to exponential decay models with decay coefficient $\Theta = 0.32$ and error $R^2 = 0.98$. }\label{fig:dists}
\end{figure}

Beyond considering the average localization length, the full distribution of localization lengths ($P(\xi_z)$) sheds further light on the fate of the exact l-bits. The histograms of the localization lengths for two representative disorder strengths at system size $L=14$ are shown in fig. \ref{fig:dists}. The distribution of localization lengths changes qualitatively with disorder strength. As shown in Figure \ref{fig:dist8_10}, at $h=10$ the peak of the distribution is below $\xi_z=1$ with the majority of  operators being localized within a single lattice spacing. The tails of the distribution show that operators with larger localization lengths are exponentially rare, $P(\xi_z) \sim \exp(-\xi_z/\Theta)$. Hence, these operators are stable to local perturbations in the Hamiltonian. The fitted value of $\Theta = 0.32$ for $h=10$.

On reducing the disorder strength to $h=4$, the peak of the distribution shifts to $\xi_z>1$. As the system approaches the critical point, a large fraction of operators have $\xi_z>1$ and the tail of the distribution can be fitted to an exponential decay $P(\xi_z) \sim \exp(-\xi_z/\Theta)$ or a power law $P(\xi_z) \sim \xi_z^{-\eta}$ equally well, as shown in fig. \ref{fig:dist4_6}. Although, there continue to exist operators with $\xi_z \approx 1$, a large part of the weight of the distribution moves to larger $\xi_z$. This shows that the l-bit operators with large localization length are more probable close to the transition in comparison to the l-bits at large disorder. %Deep in the MBL phase, the exponential rarity of l-bits with large localization lengths is crucial for their stability to perturbations on the Hamiltonian. The power law tails may render these operators unstable raising the possibility of alternative effective description of localization near the transition \cite{chandran2016higherD}.    
%\textcolor{blue}{(AP: Make log scale and log-log fitiing of the distributions in the appendix. Just to have done all that the referee suggests.)}
Interestingly, on reducing the disorder below the critical disorder strength $h_{\text{crit}}$, we find operators with $\xi_z \approx 1$ for our finite size systems. Although they form a small fraction of the distribution, it suggests that even in the `thermal' phase close to the transition there are local operators which commute exactly with the Hamiltonian. The presence of such operators is in striking contrast to the traditional understanding of non-integrable models. (Although, there are now studies of edge mode operators with very long lifetimes even in models with broken integrability \cite{kemp2017long}.) It will be interesting to study the role of these operators in the slow dynamics observed in the thermal regime close to the transition \cite{Agarwal2015Griffiths}. These are questions which are ripe for further investigation.

%Of additional interest is the decay of the tail of the distribution of localization lengths. A power law decay for the tail of the distribution is shown on Figures \ref{fig:dist4} and \ref{fig:dist6}, disorder strengths $h=4$ and $h=6$ respectively. Exponential decays are shown on Figures \ref{fig:dist8} and \ref{fig:dist10}, which at disorder strengths $h=8$ and $h=10$ are much further in the localized phase and therefore show stronger decay in the probability of highly non-local operators.

%We employed this method for systems up to $n=12$. Further, the performance of the algorithms depends heavily on the physical range of the TNS unitary transformations utilized in the method from \cite{wahl_pal_simon}; the larger the window of nontrivial action of the approximate qLIOMs from the tensor network approximation from \cite{wahl_pal_simon}, the better the eigenstate matches and the lower the localization length of the exactly constructed qLIOMs. The maximum range possible using the algorithm from \cite{wahl_pal_simon} is 8. In these simulations, we use a range of 6 due to computational constraints; our maximum possible system size is $n=12$ and the range of the unitary transformations must divide the system size. 

%\section{\label{sec:concl}Conclusions and Outlook}

\emph{Conclusions:} In this article we presented a novel technique to construct the exact l-bit operators comprising the quasi-local $\tau_i^z$ operators which commute with the Hamiltonian and $\tau_i^x$ operators which characterize the matrix elements between all the energy eigenstates. We did so by implementing the pairing structure on approximate eigenstates produced using the tensor network method from \cite{Wahl2017PRX}. We construct l-bit operators by matching the approximate eigenstates to the exact eigenstates. We find that the l-bit operators exhibit weight functions $w_i$ (see Eq. \ref{eq:weight}) that decay exponentially away from their localization centers. 

Mean localization length of $\tau_i^z$ operators increases from less than a lattice spacing at large disorder to a fraction of the system size close to the transition. This increase is consistent with a power-law divergence at $h \approx 3.5$ which is close to other exact diagonalization studies. A scaling collapse of the the data for different system sizes of the average localization length can be performed  and fitted to a power law with an exponent $\alpha \approx 0.55$ and correlation length exponent $\nu=1.11$. Both these quantities are likely to be affected by finite-size effects and may not be in the scaling regime. We find further that the distribution of localization length $\xi_z$ changes qualitatively as a function of disorder strength. At disorder strength close to the phase transition, the distribution of localization lengths shows a peak with tails which can be fitted to a power-law, where the exponent of the power changes continuously with disorder (exponential fits are also reasonable). At higher disorder strengths, the peak of the distribution shifts to lower values with exponentially decaying tails. 

Our results have several important implications for the MBL transition into the thermal phase. The power-law distribution of the length scales is consistent with the strong-disorder renormalization group studies of the MBL transition in coarse-grained phenomenological models \cite{VHA_MBLTransition, Potter:2015ab}. For finite-size systems, our treatment provides the crossover scale between thermal and the quantum critical regimes for a microscopic model \cite{Khemani2017MBLT}. The spatial structure of the operators with large localization lengths can be used to detect the backbone resonant structure which is expected to lead to thermalization. If the power-law distribution of localization lengths was to survive in the thermodynamic limit, it would suggest an intermediate phase with coexisting localized and delocalized operators. It would be interesting to search for an effective description of such a phase and its relationship to $l^*$-bit phenomenology \cite{chandran2016higherD}.   

An interesting observation is the existence of local operators at low disorder values such as $h=3$. The critical point between the MBL and thermal phases is thought to be close to $h_{\text{crit}} = 3.5$. However, our algorithm indicates the existence of local operators even for $h \leq 3.5$. This indicates that the critical disorder strength at which local operators disappear completely for system size $L=14$ is below $h=3$.  It could indicate that a different effective model is required to describe the localized phase near the phase transition.

Note added: After completion of this work, we became aware of the independent similar work presented in Refs. \cite{goihl2017construction, thomson2017Flow}. \emph{Acknowledgments:} We would like to thank Fabian Essler and Paul Fendley for helpful discussions. S.H.S. and T.B.W. are both supported by TOPNES, EPSRC grant number EP/I031014/1. S.H.S. is also supported by EPSRC grant EP/N01930X/1. This project has received funding from the European Union’s Horizon 2020 research and innovation programme under the Marie Skłodowska-Curie grant agreement No. 749150. The work of A.P. was performed in part at the Aspen Center for Physics, which is supported by National Science Foundation grant PHY-1066293. Statement of compliance with EPSRC policy framework on research data: This publication is theoretical work that does not require supporting research data.

\bibliographystyle{unsrtnat}
\bibliography{biblioMBL}

\begin{thebibliography}{40}
\bibitem[FMB()]{FMBLfootnote}
For the purposes of this work, we'll always consider $\text{FMBL}$ systems and
  for the sake of brevity use the term $\text{MBL}$. $\text{A}$lthough, it is
  important to keep in mind that the phenomena of $\text{MBL}$ can be more
  general.
\end{thebibliography}

\appendix

\section{Details of construction of exact l-bit operators}

The tensor network approach described in \cite{Wahl2017PRX} provides an efficient approximation of all eigenstates of MBL systems. This method employs a 2-layer ansatz comprising local unitary rotations to transform trivial spin states to approximate MBL eigenstates. When stitched together, these layers of local unitary rotations create a unitary matrix that approximately diagonalizes the Hamiltonian of the system. %In \cite{pollmann_khemani_cirac_sondhi_2016}, the physical range of these transformations is limited by the number of layers used; each additional layer increases computational cost exponentially. In \cite{wahl_pal_simon}, the physical range of the transformation is extended by using larger local unitary rotations. 
This algorithm produces operators with limited support on the lattice; an l-bit operator produced by these methods has nontrivial support on a finite region and acts trivially everywhere else. However, it produces rotations that preserve local structure, so it provides a pairing structure over approximate eigenstates which can be used to construct exact l-bit operators.

Our goal is to craft $\ket{\alpha},$ $\ket{\beta_{i,\alpha}}$ pairings using the approximations given by the tensor network approach. The tensor network approximation yields a unitary matrix whose columns are approximate eigenstates $\ket{a}$ of the Hamiltonian. However, this unitary matrix does not exactly diagonalize the Hamiltonian. The columns of this unitary matrix encode a pairing structure on the approximate eigenstates, which we label $\ket{a}$, $\ket{b_{i,a}}$. This built-in pairing structure is given by the indexing of the unitary matrix. To impose this pairing structure on the exact eigenstates of a system, we attempt to find a one-to-one mapping that matches approximate eigenstates to exact eigenstates. A proper mapping will produce l-bit operators with quasi-local action and exact commutation with the Hamiltonian.

The purpose of this mapping procedure, which we call \textit{matching}, is to assign each exact eigenstate a place in the l-bit spin structure. If an exact eigenstate $\ket{\alpha}$ is matched to an approximate eigenstate $\ket{a}$, with l-bit assignment $\{+ - + + - +\}$ for example, then the exact eigenstate is assigned the same l-bit label. A one-to-one mapping assigns each exact eigenstate to each position in the spin structure on MBL eigenstates.

\begin{figure}
\resizebox{\linewidth} {!}  {
\begin{tikzpicture}
\node (a) at (0,0) {\large $\ket{m}$} node[below = .1 of a] {\tiny $\{++++\}$};
\node (b) at (2,1.5) {\large $\ket{\text{l}_1}$};
\node (c) at (2,0) {\large $\ket{\text{l}_2}$};
\node (d) at (2,-1.5) {\large $\ket{\text{l}_3}$};
\node (e) at (2.5,1.5) { $\leftrightarrow$} node[below = .1 of e] {\tiny $\{-+++\}$};
\node (f) at (2.5,0) { $\leftrightarrow$} node[below = .1 of f] {\tiny $\{+-++\}$};
\node (g) at (2.5,-1.5) { $\leftrightarrow$} node[below = .1 of g] {\tiny $\{+++-\}$};
\node (h) at (3,1.5) {\large $\ket{\lambda_1}$};
\node (i) at (3,0) {\large $\ket{\lambda_2}$};
\node (j) at (3,-1.5) {\large $\ket{\lambda_3}$};
\node (k) at (5,0) {\large $\ket{\mu}$};
\node[] (l) at (4,-2.75) {\large $\Rightarrow \ket{\text{m}} \leftrightarrow \ket{\mu}$} node[below = .05 of l] (m) {\tiny $\{++++\}$};
\node [draw=blue, fit= (l) (m)] {};
\draw[red,->] (a) -- (b) node[black,pos=.6,above,sloped] {$\tau^{x,\text{TN}}_1$};
\draw[red,->] (a) -- (c) node[black,pos=.7,above,sloped] { $\tau^{x,\text{TN}}_2$};
\draw[red,->] (a) -- (d) node[black,pos=.6,above,sloped] { $\tau^{x,\text{TN}}_4$};
\draw[red,dashed,->] (h) -- (k) node[black,pos=.5,above,sloped] {$\tau^{x,\text{TN}}_1$};
\draw[red,dashed,->] (i) -- (k) node[black,pos=.4,above,sloped] {$\tau^{x,\text{TN}}_2$};
\draw[red,dashed,->] (j) -- (k) node[black,pos=.5,above,sloped] {$\tau^{x,\text{TN}}_4$};
\end{tikzpicture}
}
\caption[\linewidth]{Figure depicting a matching algorithm between tensor network approximate states and exact eigenstates. The set of approximate eigenstates has an existing pseudo-spin structure. Given a set of already matched approximate and exact states, $\ket{\text{a}_i}$ and $\ket{\alpha_i}$ respectively, an approximate eigenstate $\ket{\text{b}}$ can be matched to an exact eigenstate $\ket{\beta}$ if the same transformation that exactly takes $\ket{\text{b}}$ to $\ket{\text{a}_i}$ (solid arrows) roughly takes $\ket{\alpha_i}$ to $\ket{\beta}$ (dashed arrows). This procedure takes advantage of the fact that ${\tau^x_i}^2 = 1$}
\label{fig:matching}
\end{figure}
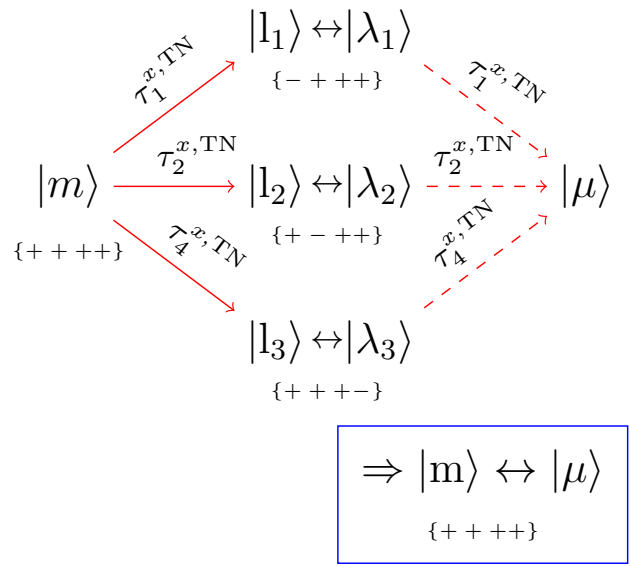

Matching approximate and exact eigenstates entails two parts: first finding approximate eigenstates and exact eigenstates with an inner product close to $1$, and second matching the remaining eigenstates using an algorithm described below and in figure \ref{fig:matching}. Oftentimes, especially deep in the localized phase, many pairings are obvious as overlap between approximate eigenstates and exact eigenstates is high. In these cases, one can simply carry out the first part of the matching algorithm by finding the best matched approximate eigenstate for each exact eigenstate using the inner product and pairing them together. Closer to the phase transition, matchings are less obvious, requiring the use of the second part of the algorithm.

For the first part of the matching process, we match any eigenstates for which $\left| \bra{\alpha}\ket{a} \right| > t$, where $t$ is some threshold. For our calculations, we set $t = 0.6$. Increasing the threshold increases computational cost, while decreasing it runs the risk of making poor assignments. When a match above the threshold is found, we match $\ket{a}$ to $\ket{\alpha}$ and assign $\ket{\alpha}$ the same l-bit label as $\ket{a}$, meaning that ${l^a_i} = {l^\alpha_i}$ and the set $\left\lbrace \ket{b_{i,a}} \right\rbrace$ corresponds to the set $\left\lbrace \ket{\beta_{i,\alpha}} \right\rbrace$. 

Because some eigenstates cannot be matched above this threshold, we utilize the second part of the matching process, using the already-matched states to inform the new matches. Consider $\{\ket{\lambda_k}\}$ the set of exact eigenstates matched to a tensor network approximate eigenstate $\{\ket{l_k}\}$ ($\ket{\lambda_1}$ matches $\ket{l_1}$ and so on). There is left a set of unmatched exact states $\{\ket{\mu}\}$ and unmatched tensor network states $\{\ket{m}\}$. 

We can now use the pairing structure on approximate eigenstates to inform new matches. An unmatched approximate eigenstate $\ket{m}$ has a set of $j$-partners in $\{\ket{l_k}\}$ which each have matches in $\{\ket{\lambda_m}\}$. Conveniently, the approximate $\tau_i^x$ operators produced from the tensor network algorithm give us the transformations from $\ket{m}$ to its $j$-partners in the already matched set $\{\ket{l_{j,m}}\}$. If the same transformations roughly take the exact eigenstate matches of $\{\ket{l_{j,m}}\}$, labeled $\{\ket{\lambda_{j,m}}\},$ to an unmatched exact eigenstate $\ket{\mu}$, then we match $\ket{\mu}$ and $\ket{m}$. To make assignments iteratively, we find these new matches one at a time by scanning over $\{\ket{\mu}\}$ and $\{\ket{m}\}$ to find the eigenstates from each eigenstate that maximally fit this pattern. Once a match is made, it can be used to inform new matches in new iterations. Figure \ref{fig:matching} illustrates this relation.

To this end, we find the eigenstates in $\{\ket{\mu}\}$ and $\{\ket{m}\}$ that maximize the function
\begin{equation} \label{eq:flow}
f(\mu,m) = \sum_{i \in s(m)} | \bra{\lambda_{i,m}} \tau^{x,\text{TN}}_i \ket{\mu} |^2 
\end{equation}
where $\tau^{x,\text{TN}}_i$ is the l-bit operator yielded from the tensor network approximation; $s(m)$ is the set of sites $j$ where the $j$-partner of $\ket{m}$ is in $\{\ket{l_k}\}$; and $\ket{\lambda_{i,m}}$ is the exact eigenstate matched to the the $i$-partner of $\ket{m}$. %We match an exact state $\ket{\mu}$ to an approximate state $\ket{m}$ if the tensor network operator $\tau^{x,\text{TN}}$ taking $\ket{\alpha_m}$ to $\ket{\beta}$ also approximately takes $\ket{a_m}$ to $\ket{b}$. 

After finding the maximizing values of $ \ket{\mu}$ and $ \ket{m}$ for Equation \ref{eq:flow}, we match these two states, add them to the set of matched states and iterate until all matches have been made. At larger system sizes and for lower disorder strengths, the number of unmatched states can make this process computationally expensive. As such, making more than one assignment on each iteration expedites the process.

After a complete set of pairings is made, operators are constructed as described in equations (2)-(4) in the main text. %(\ref{eq:lbitsx})-(\ref{eq:lbitsz}).

The phase of the eigenstates produced by exact diagonalization presents another consideration. The eigenstates produced by exact diagonalization are allowed an arbitrary scalar phase. Because our Hamiltonian is real and can therefore be diagonalized by an orthogonal matrix, MATLAB produces eigenstates with arbitrary sign. However, the sign of the eigenstates affects the calculation of $\tau^x_i$ and $\tau^y_i$ as shown in equations (2) and (3) (in the main text), %(\ref{eq:lbitsx}) and (\ref{eq:lbitsy})%
requiring us to choose the `correct' sign in order to construct l-bit operators. Note that the calculation of $\tau^z_i$ as shown in Equation (4) (in the main text) %(\ref{eq:lbitsz}) 
is unaffected by the sign.

To choose the sign of an eigenstate after the matching process is completed, we use the following algorithm: For an eigenstate and its $i$-partner $\ket{\alpha}$ and $\ket{\beta_{i,\alpha}}$, we take $\hat{O} = \Tr_{\bar{i}} [\ket{\alpha} \bra{\beta_{i,\alpha}} + \ket{\beta_{i,\alpha}} \bra{\alpha}]$, where $\Tr_{\bar{i}}$ is a partial trace over all sites except for $i$. If $\hat{O}$ resembles $\sigma_x$, the signs of both eigenstates remain unchanged. If $\hat{O}$ resembles $-\sigma_x$, then the sign of $\ket{\beta_{i,\alpha}}$ is flipped. The sign of $\ket{\beta_{i,\alpha}}$ is then set and the process is repeated until all eigenstates have been assigned a sign. After this process is completed $\hat{O}$ should resemble $\sigma_x$ for any pair $\ket{\alpha}$ and $\ket{\beta_{i,\alpha}}$, however the assignment of sign could not be carried out for all pairs. Although, this doesn't change the algebraic properties of the operators, it does effect their localization lengths.

\section{Localization length fitting procedure}

As described in the main text, localization lengths are determined by fitting an exponential decay on the weight function of the operator. This process is explained in more detail below.

Consider an operator acting on site $i$. The weight function operator should peak on site $i$ and decay on either side. As an ansatz, we assume the form of this decay to be exponential in nature, meaning that the weight function resembles $w(j) \propto e^{- \left| i-j \right|/\xi},$ where $i$ is the site of peak action of the operator. We consider the points on the weight function from $i$ to the furthest edge. Taking the natural logarithm of the weight at each site minus the weight at the edge of the system, we then make a linear fit on each point. The absolute value of the slope of this line is $1/\xi_b$, where $\xi_b$ is the best fitting $\xi$ for the ansatz described above. We take the localization length $\xi$ to be the value $\xi_b$.

This ansatz turns out to be accurate in practice according to the mean squared error of the linear fit. Deep in the localized phase at $h=10$ for $L=14$, the linear fits have an average $R^2 = .85$. Moving into the thermal phase, the average $R^2$ remains relatively high, indicating that some operators retain exponential decay characteristics. At $h=6$, the average $R^2 = .84$. At $h=4$, the average $R^2 = .86$. At $h=1$ the average $R^2 =  .80$.

\section{Distribution of l-bits}

\begin{figure}
	\begin{minipage}{0.8\linewidth}
		%\begin{subfigure}
		\includegraphics[width=0.85\linewidth]{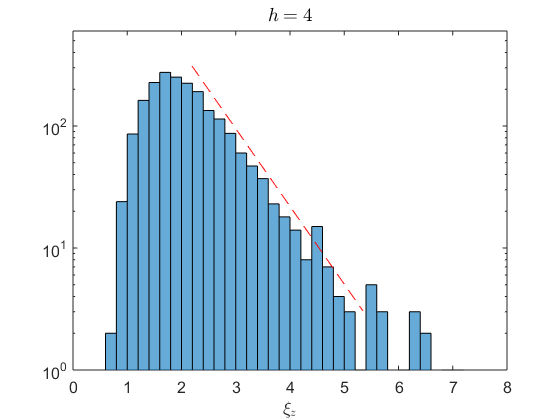}
		\subcaption{\label{fig:dist_4_exp}}
		%\caption{\label{fig:dist4_6}}
		%\end{subfigure}
	\end{minipage}
	\begin{minipage}{0.8\linewidth}
		%\begin{subfigure}
		\includegraphics[width=0.85\linewidth]{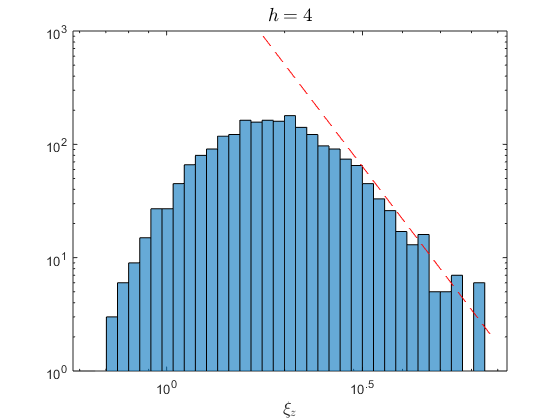}
		\subcaption{\label{fig:dist_4_pow}}
		%\caption{\label{fig:dist4_6}}
		%\end{subfigure}
	\end{minipage}
	%\begin{subfigure}
	\begin{minipage}{0.8\linewidth}
		\includegraphics[width=0.85\linewidth]{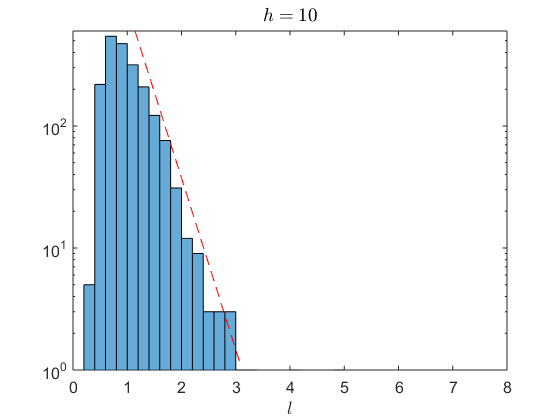}
		\subcaption{\label{fig:dist_10_exp}}
		%\caption{\label{fig:dist8_10}}
	\end{minipage}
	%\label{fig:dist8_10}
	%\end{subfigure}
	\caption[width=\linewidth]{Histograms of localization lengths of $\tau_i^z$ at system size $L=14$ on log-log and semi-log plots. The linear decays of the tails of the semi-log plots (fig. \ref{fig:dist_4_exp} for $h=4$ and fig. \ref{fig:dist_10_exp} for $h=10$) indicate an exponentially decaying tail. The linear decay of the tail on the log-log plot (fig. \ref{fig:dist_4_pow} for $h=4$) indicates a power-law decay. Note that in this plot, the histogram has been re-binned to show equal bin size on a log scale. }\label{fig:dists2}
\end{figure} 
Additional figures for the distribution of l-bit localization lengths are given in fig. \ref{fig:dists2}. The plots show the same data given in the main text, but on log-log and semi-log plots to exhibit the power-law and exponential decay fits found in the tails. In the main text, we concluded that fits of approximately equal quality could be found for the distribution of localization lengths at $h=4$ using both exponential decay and power-law fits. This is exhibited by the fact that the tails show a linear decay on the log-log (fig. \ref{fig:dist_4_exp}) and semi-log (fig. \ref{fig:dist_4_pow}) plots. We additionally concluded that the distribution of localization lengths at $h=10$ exhibits an exponential law decay, exhibited by the linear nature of the decay on a semi-log plot (fig. \ref{fig:dist_10_exp}).

\end{document}